# Neutron scattering by ultrasound in mosaic single crystals

*E. Iolin (1,2), B. Farago (3), F. Mezei (4, 5), L. Rusevich (1), J. Pieper (4,6),
A. Buchsteiner (4), G. Ehlers (7), W. Haussler (8)*

[1] Institute of Physical Energetic, Riga, Latvia
[2] Physics Department, Lehigh University, Bethlehem, PA, USA
[3] Institute Laue - Langevin, Grenoble, France
[4] Hahn - Meitner Institute, Berlin, Germany
[5] ESS AB, Lund, Sweden
[6] Institute of Physics, University of Tartu, Tartu, Estonia
[7] Oak-Ridge National Laboratory, SNS, Oak Ridge, TN, USA
[8] ZWE FRM-II, Garching, Germany

**ABSTRACT**

The Time-of-Flight and Neutron-Spin-Echo techniques have been applied for the research of ultrasonic effect at the cold neutron diffraction in mosaic single crystals (KBr and mica). It was found that h.f. ultrasound intensifies quasi-elastic neutron scattering and this effect cannot be explained as a result of neutron scattering by thermal phonons in a sample heated by ultrasound; neither can it be interpreted in the context of standard secondary extinction theory. A new modified model is proposed based on the assumption that the angular distribution function describing mosaic block disorientation is not smooth but contains deep narrow troughs. This model is qualitatively equivalent to the preposition about existence large block structure of sample. The corresponding kinetic equations are formulated and exactly solved. The results obtained correlate well with experimental data. Our results evidence that such an approach is applicable for studying the ultrasonic effect at neutron diffraction in different technologies dealing with mosaic single crystals.

# 1. Introduction.

Neutron and X-ray scattering by a high-frequency (h.f.) ultrasonic acoustic wave (AW) in a perfect single crystal (s.c.) have been studied for a long time [1]. As a rule, dependence of the intensity of quasi-elastic scattering, $Iht$, is studied as a function of the voltage across the transducer, the a.c. frequency, the acoustic wave polarization type (LAW or TAW), etc. Ordinary term 'quasielastic scattering intensity', $Iht$, means the scattering around Bragg angle after deduction scattering at the thermal phonons. We use the same standard definition of $Iht$. The energy and momentum of ultrasonic phonons are very small. Therefore value of $Iht$ contains contributions from zero, $\pm 1$, $\pm 2$, $\pm\ldots$ ultrasonic phonon scattering. Ultrasound strongly increases $Iht$ in the case of perfect silicon s.c. samples (see e.g. [2]). It was also found that new Pendellosung (interference) beatings with unusually large extinction length are created by the weak ultrasonic wave [3 - 5]. Such X-ray and neutron beatings are sensitive to very small deformations in silicon. A new unusual effect was observed for the case of a deformed Si plate: it turns out that low amplitude AW *decreases* the intensity of neutron scattering approximately by half [6, 7]. These neutron and X-ray data are in agreement with the dynamical theory calculations (partly they were even predicted [3, 8]). Experimental results [9 - 14] concerning inelastic neutron scattering by ultrasound in silicon s.c. also agree with theory.

The situation with studying AW effects differs for the case of neutron and X-ray scattering by mosaic single crystals, for which the theory of ultrasonic effect is almost absent and experimental works are few in number. In order to observe the measurable effects it is necessary to use more or less strong AW, which entails heating of the sample and additional undesirable scattering by the arising thermal phonons.

This paper is concentrated on the experimental and theoretical research into ultrasonic effects at the neutron scattering in mosaic s.c. Sixteen years ago [9 - 14] we began to study the inelastic neutron scattering by ultrasonic AW. We used the Neutron-Spin-Echo (NSE) technique and observed, for the first time, zero (elastic), $Ih(0)$, one, $Ih(1)$, double, $Ih(2)$, etc. ultrasonic phonon scattering. We applied the NSE technique for the case of mosaic single crystal samples and obtained more detailed information about ultrasonic effects than it had been possible before, studying quasi-elastic scattering. However, we faced the problem that it was difficult to carry out NSE experiments using different neutron wavelength, $\lambda$. That is why recently we have initiated ultrasonic AW experiments based on the Time-of-Flight (ToF) technique, using the V3 instrument at the Berlin Neutron Scattering Center (BENSC). Of course, it is impossible to observe inelastic neutron scattering by ultrasound using such an instrument. However, it is easy to derive quasi-elastic scattering intensity $Iht$ from the total intensity of scattering and to exclude neutron scattering by thermal phonons in ToF experiments. Here we show first results of our ToF ultrasonic studies for the case of mosaic s.c. and give an analysis of our earlier NSE results. Our experiments were done for the case of cold neutron scattering. The theory is essentially simplified in this case.

The theoretical description of this effect is proposed in Sect. 2 of this paper. It is based on the observation that exchange of energy between a cold (slow) neutron and the ultrasound leads to the existence of a separate inelastic channel of scattering in each mosaic block. The corresponding kinetic equations that take into account all ultrasonic transitions are formulated and exactly solved for the case of an arbitrary angular distribution function, *F*, describing mosaic block orientation, and coherent and incoherent ultrasonic waves. The Fourier analysis of calculated NSE signal allows for finding the intensity of quasi-elastic, elastic, one- and double-phonon scattering. We found that we can understand our experimental data only if we supposed that the function, *F*, describing mosaic block disorientation in the plane of scattering is not a smooth (e.g. gaussian) function but it contains deep narrow troughs. Neutrons "corresponding" to the troughs are not scattered if ultrasound is removed. These neutrons are inelastic scattered when ultrasound is applied.

The results are compared with our experimental data (see Sect. 3). As far as we know it is the first more or less successful attempt to explain ultrasonic effect at the diffraction in the mosaic s.c. The observed and the calculated ultrasonic effects at the diffraction are found to strongly depend on the parameters of the function of mosaicity distribution. Therefore the intensity of quasi-elastic scattering and the spectrum of ultrasonic effect at the diffraction can be even considered (in general) as some test of mosaic structure "perfection".

General discussion of the results and potentialities hidden in our approach are found in the Summary Sect. Here it could be mentioned that the energy resolution achieved in our NSE experiments with ultrasound is much better than that for standard NSE technique. Below we will explain the reason for this.

## 2. Theory.

### 2.1 *General approach*

We will suppose that the ultrasonic wavelength, $\lambda_S$, is much greater than the mosaic block linear dimension, *L*, that is $\lambda_S \gg L$. Each mosaic block is displaced as a whole by the acoustic wave with amplitude **W**, wave vector $\mathbf{k}_S$ and angular frequency $\omega_S$. Displacement **U** for the case of a traveling acoustic wave is written in the form:

$$\mathbf{U}_S = \mathbf{W}\sin(\omega_S t - \mathbf{k}_S \mathbf{r}) \tag{1}$$

The neutron momentum is changed by value $\boldsymbol{\delta}$ due to the absorption (emission) of the ultrasonic phonon as

$$\boldsymbol{\delta} = \mathbf{k}_S \pm \boldsymbol{\delta}_D, \quad k_S = \omega_S/V_S, \quad \boldsymbol{\delta}_D = \mathbf{e}_d \omega_S/(V_N \cos\theta_B), \quad \mathbf{e}_d = (\mathbf{k}_0 + \mathbf{k}_h)/|\mathbf{k}_0 + \mathbf{k}_h|,$$
$$\mathbf{k}_h = \mathbf{k}_0 + \mathbf{G}, \quad (\boldsymbol{\delta}_D \mathbf{G}) = 0, \quad k_h = k_0, \quad 2k_0\sin(\theta_B) = G \tag{2}$$

Here $\mathbf{k}_0$, $\mathbf{k}_h$ are the wave vectors of incident and elastically diffracted beam, respectively; **G** is the vector of scattering; $\theta_B$ is the Bragg angle. $V_S$ and $V_N$ are the velocities of ultrasonic AW and neutron, respectively. Further we will consider the case of cold (slow) neutrons with only the second (Doppler's) term in (2), $\boldsymbol{\delta}_D$, significant. For example, parameter $\delta_D/k_S$ is equal to the value of 4.3 and 11 for the case of transverse and longitudinal acoustic wave, respectively, near the reflection (200) at the crystal KBr (neutron wavelength 5.85Å). It is interesting that vector $\boldsymbol{\delta}_D$ is parallel to the reflecting planes. The importance of this circumstance will be seen in the following. We will neglect term $k_S$ in the expression (2), This approximation is really exact when the ultrasound is spreading perpendicularly to the plane of scattering. We will restrict our consideration to this case. The amplitude of neutron scattering by a mosaic block is $f(t) \sim exp(iGW sin(\omega_S t - k_S r))$.

We will distinguish two limiting cases:

1) A mosaic block of small size, $L < 1/\delta_D$. In this case absorption (emission) of the AW phonon does not lead to a visible diffusion of a neutron towards the side of the reflection curve; AW simply transforms the elastic scattering to an inelastic, and the total intensity of quasi-elastic scattering, $Iht$, is independent of ultrasonic excitation.

2) A mosaic block of medium size, $\delta_D L > 1$ (but $L < \tau$, $\tau$ is the neutron extinction length[1]). The elastic and inelastic scattering are realized in different parts of the momentum space of one and the same mosaic block. Therefore the channels of elastic and inelastic scattering should be considered as separated. In this case the final effect at the diffraction depends on the value of angular disorientation of the mosaic blocks. Multiply scattering inside a block is very small.

We have done calculations for the second case only. It can be shown that the case of small mosaic blocks could be considered as limiting in the corresponding expressions when there are no deep troughs in the distribution function of mosaic disorientation.

We will begin with analyzing the symmetrical Bragg scattering by a mosaic s.c. plate (Fig.1).

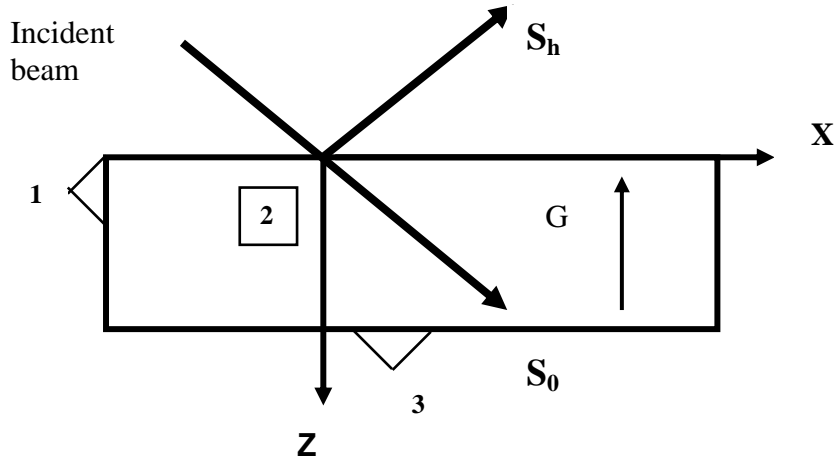

Fig.1. Geometry of diffraction. 1, 2 — TAW, 3 — LAW piezoelectric $LiNbO_3$ transducers.

We suppose, as usual [16] that the phases of neutron scattering amplitude by different mosaic blocks do not correlate. For the simplicity we limit ourselves by the case when ultrasonic wave amplitude is uniform over the sample (although the general analysis and calculation can also be performed without this simplification).
We take into account all ultrasonic transitions.

---

[1] We should consider dynamical scattering if $\tau \gg L$.

Kinetic equations for neutrons are written as following:

$$\frac{\partial n_0(P_0, N)}{\partial S_0} = -\sum_{m=-\infty}^{m=\infty} aJ_m^2(\mathbf{GW})F(-m\Omega_S/2 - P_0)n_0(P_0, N) +$$

$$+ \sum_{m=-\infty}^{m=\infty} aJ_m^2(\mathbf{GW})F(m\Omega_S/2 - P_0)n_h(-P_0, N+m),$$

$$\frac{\partial n_h(-P_0, N)}{\partial S_h} = -\sum_{m=-\infty}^{m=\infty} aJ_m^2(\mathbf{GW})F(m\Omega_S/2 - P_0)n_h(-P_0, N)$$

$$+ \sum_{m=-\infty}^{m=\infty} aJ_m^2(\mathbf{GW})F(-m\Omega_S/2 - P_0)n_0(P_0, N+m) \quad (3)$$

$n_0(P_0,N)$, $n_h(-P_0,N)$ is the number of neutrons in the incident and the diffracted beam, respectively; $m$ is the number of absorbed ultrasonic phonons, $m=0,\pm1,\ldots$; $F$ is the function describing an *arbitrary* angular distribution of the mosaic block orientation in the plane of scattering.; $J_m$ is the Bessel function [1]; $a$ is a numerical factor (defined below). Equations (3) are written in the dimensionless form.

$$x \Rightarrow x\tau/\pi, \quad z \Rightarrow z\tau/\pi * \text{tg}\Theta_B,$$
$$S_0 = x + z, \quad S_h = x - z,$$
$$\Omega_S = \omega_S \tau/(\pi V_n \cos\Theta_B) = \delta_D \tau/\pi \quad (4)$$

$\tau$ is the extinction length; $P_0$ is the momentum – deviation from the exact Bragg condition (Lorentz point) along direction of the diffracted beam, $S_h$ (Fig.1). Value of $P_0=\pm m\Omega_S/2$ is corresponding position of $\pm m$-th ultrasonic phonon satellite.

The boundary condition for the case of symmetrical Bragg's reflection by a plate with thickness $T$ may be written in the form:

$$n_0(P_0, N)\big|_{Z=0} = \delta_{N,0}/2\pi, \quad n_h(-P_0, N)\big|_{Z=T} = 0 \quad (5)$$

Generally, equations (3) are similar to the standard secondary extinction equations [16]. The main difference is that the ultrasonic transitions and corresponding momentum change are included in (3).

It is convenient to solve (3) using Fourier transform over the number of phonons:

$$n_0(P_0, N) = \int_{-\infty}^{\infty} n_0(P_0, \xi)\exp(i\xi N)d\xi, \quad n_h(-P_0, N) = \int_{-\infty}^{\infty} n_h(-P_0, \xi)\exp(i\xi N)d\xi \quad (6)$$

$$(\frac{\partial}{\partial x} + \frac{\partial}{\partial z})n_0(P_0,\xi) = -A(P_0)n_0(P_0,\xi) + B(P_0,\xi)n_h(-P_0,\xi) \qquad (7a)$$

$$(\frac{\partial}{\partial x} - \frac{\partial}{\partial z})n_h(-P_0,\xi) = -A(P_0)n_h(-P_0,\xi) + B^*(P_0,\xi)n_0(P_0,\xi) \qquad (7b)$$

$$A(P_0) = \sum_{m=-\infty}^{m=\infty} a 2 J_m^2(\mathbf{GW}) F(-m\Omega_S/2 - P_0) \qquad (7c)$$

$$B(P_0,\xi) = \sum_{m=-\infty}^{m=\infty} a 2 J_m^2(\mathbf{GW}) F(m\Omega_S/2 - P_0)\exp(im\xi), \quad A(P_0) \geq |B(P_0,\xi)| \qquad (7d)$$

$$n_0(P_0,\xi)|_{Z=0} = 1/4\pi^2, \quad n_h(-P_0,\xi)|_{Z=T} = 0 \qquad (8)$$

We are interested in the diffraction of wide beams. Therefore operator $\partial/\partial x$ could be omitted in (7) and the solution of (7), (8) can be written in a simple form. Fourier's component of the diffracted beam intensity will be

$$n_h(-P_0,\xi)|_{Z=0} = \frac{1}{4\pi^2} \frac{\exp(-i\varphi)\sin(\rho)}{1+\cos(\rho)cth(AT\cos(\rho))}, \quad 0 \leq \rho < \pi/2, \quad -\pi \leq \varphi < \pi$$

$$B(P_0,\xi) = |B(P_0,\xi)|\exp(i\varphi), \quad |B(P_0,\xi)| = A(P_0)\sin(\rho), \qquad (9)$$

The total intensity of quasi-elastic scattering is

$$I_{htB} = \int_{-\infty}^{\infty} \sum_{N=-\infty}^{N=\infty} n_h(-P_0,N)|_{Z=0} dP_0 = \frac{1}{2\pi} \int_{-\infty}^{\infty} \frac{AT}{1+AT} dP_0 \qquad (10)$$

*Non-normalized* Neutron-Spin-Echo (NSE) signal is expressed as

$$S_B(t) \equiv \int_{-\infty}^{\infty} \sum_{N=-\infty}^{N=\infty} n_h(-P_0,N)|_{Z=0} \cos(N\omega_S t) dP_0 = \frac{1}{2\pi} \int_{-\infty}^{\infty} \frac{\cos(\varphi)\sin(\rho)}{1+\cos(\rho)cth(AT\cos(\rho))}\bigg|_{\xi=\omega_S t} dP_0 \qquad (11)$$

Here $t$ is the NSE time, which is proportional to the value of magnetic field in the NSE spectrometer coils [17], [18].

Similar calculations may be done for the case of symmetrical Laue scattering. We have:

$$n_{hL}(-P_0,\xi) = \frac{1}{4\pi^2}\exp(-i\varphi)\exp(-AT)sh(|B|T) \qquad (12)$$

The total intensity of quasi-elastic scattering is described by the expression:

$$I_{htL} = \frac{1}{4\pi} \int_{-\infty}^{\infty} (1-\exp(-2AT))dP_0 \qquad (13)$$

The non-normalized NSE signal is

$$S_L(t) = \frac{1}{2\pi} \int_{-\infty}^{\infty} \cos(\varphi)\exp(-AT)sh(|B|T)\bigg|_{\xi=\omega_S t} dP_0 \qquad (14)$$

Equations (3) contain so far unspecified parameter $a$. This parameter can be found by comparison (3) with corresponding equations for the neutron scattering in a mosaic single crystal (see [16]) with gaussian distribution of the mosaic block angular orientation. We thus obtain

$$a = 4\pi t g(\theta_B), \quad F_G(P_0) = \frac{\lambda}{\eta\tau\sqrt{2\pi}} \exp(-0.5(\frac{P_0\lambda}{\eta\tau})^2) \tag{15}$$

where $\tau$ is the extinction length and $\eta$ is the mosaicity.

We supposed above that the ultrasonic acoustic field is "ordinary", i.e. coherent. This approximation is correct when the ultrasonic transducer excites only one mode in a sample. However, as a rule there are many modes simultaneously excited in our experiments, since the h.f. ultrasonic wavelength $\lambda_S \sim 30-100$ μm is much smaller than the linear size of samples. The problem of coherent and incoherent ultrasound has been discussed in literature since the paper by Ruby and Bolef [19] concerning experimental research of Mossbauer's effect with a radioactive source ultrasonically excited with a frequency of 20 MHz. It was predicted theoretically that the intensity of Mossbauer's zero phonon line is $I_M \sim J_0^2(\mathbf{K}_\gamma \mathbf{W})$, where $K_\gamma$ is the γ- quantum momentum. Therefore $I_M$ should oscillate as a function of ultrasonic wave amplitude $W$. However no oscillations were observed in experiment [19]: $I_M$ was simply decreasing at the higher values of the ultrasonic wave amplitude. Ruby and Bolef treated this as evidence that "perhaps 50% of the $Fe^{57}$ nuclei were moving much more slowly than the remainder". Abragam [20] interpreted the observations of [19] as evidence that incoherent ultrasonic waves were excited in the sample. This incoherent ultrasonic field is qualitatively similar to the effective "heating" of phonons with ultrasonic frequency. Abragam inference corresponds to the model of a fully chaotic ultrasonic field and the gaussian amplitude distribution function

$$f(w) = 2\exp(-w^2/w_0^2)w/w_0^2, \quad w_0^2 = <w^2> \equiv \int_0^\infty f(w)w^2 dw \quad ,$$

so that the intensity of the Mossbauer line, $<I_M>$, is

$$<I_M> \sim \int_0^\infty J_0^2(\mathbf{K}_\gamma \mathbf{W})\exp(-\frac{(\mathbf{K}_\gamma \mathbf{W})^2}{<(\mathbf{K}_\gamma \mathbf{W})^2>})\frac{d|\mathbf{K}_\gamma \mathbf{W}|^2}{<(\mathbf{K}_\gamma \mathbf{W})^2>} =$$
$$\exp(-<(\mathbf{K}_\gamma \mathbf{W})^2>/2)I_0(<(\mathbf{K}_\gamma \mathbf{W})^2>/2) \tag{16}$$

Here $I_m$ is a modified m-order Bessel function and the above mentioned oscillations are absent in an according to the experiment [19].

We will apply the Abragam approach, having simply replaced the coefficients $J_m^2(GW)$ of the m-phonon transition probability by their averaged values $<J_m^2(GW)>$ in the kinetic eq-s (3):

$$<J_m^2(\mathbf{GW})> = \exp(-<(\mathbf{GW})^2>/2)I_m(<(\mathbf{GW})^2>/2) \tag{17}$$

Here $<(GW)^2>$ is the average value of $(GW)^2$.

For the case of incoherent ultrasonic waves the intensity of quasi-elastic scattering and the non-normalized NSE signal are written in a simple form:

$$<I_{htB}> = \frac{1}{2\pi}\int_{-\infty}^{\infty}\frac{<A>T}{1+<A>T}dP_0, \quad <A> = \sum_{m=-\infty}^{m=\infty} a2<J_m^2(\mathbf{GW})>F(-m\Omega_S/2 - P_0) \quad (18)$$

$$<S_B(t)> = \frac{1}{2\pi}\int_{-\infty}^{\infty}\frac{\cos(<\varphi>)\sin(<\rho>)}{1+\cos(<\rho>)cth(<A>T\cos(<\rho>))}\bigg|_{\xi=\omega_S t}dP_0$$

$$<B> = \sum_{m=-\infty}^{m=\infty} a2<J_m^2(\mathbf{GW})>F(m\Omega_S/2 - P_0)\exp(im\xi), \quad 0 \leq<\rho><\pi/2, \quad -\pi\leq<\varphi><\pi$$

$$<B> = |<B>|\exp(i<\varphi>), \quad |<B>| = <A>\sin(<\rho>), \quad (19)$$

$$<I_{htL}> = \frac{1}{4\pi}\int_{-\infty}^{\infty}(1-\exp(-2<A>T))dP_0,$$

$$<S_L(t)> = \frac{1}{2\pi}\int_{-\infty}^{\infty}\cos(<\varphi>)\exp(-<A>T)sh(|<B>|T)\bigg|_{\xi=\omega_S t}dP_0 \quad (20)$$

Kinetic equations can be essentially simplified for the case of small-sized mosaic blocks. We should omit the $m\Omega_S/2$ term from function $F$ (see (3)–(7)). Corresponding results will be similar to (6)–(20). For the case of *coherent* ultrasound we will have

$$A(P_0) = 2aF(-P_0), \quad B(P_0,\xi) = 2aF(-P_0)J_0(2GW\sin(\omega_s\xi/2)) \quad (21)$$

and for *incoherent* ultrasound

$$A(P_0) = 2aF(-P_0), \quad B(P_0,\xi) = 2aF(-P_0)\exp(-<(\mathbf{GW})^2>\sin(\omega_s\xi/2)^2) \quad (22)$$

## *2. 2 Analysis of the kinetic equation solution.*

### 2.2.1 Medium-sized mosaic blocks. The effect of the coherent ultrasonic wave.

We limit ourselves by the case of Bragg geometry (results for the case of Laue geometry are similar to ones). For our analysis we will take three extreme cases.

**1).** A thin crystal, $AT<<1$. In this case the probability of the multiple scattering is very low. Ultrasonic effect on the intensity of quasi-elastic scattering is absent. The NSE signal is calculated as

$$S_B(t) = S_L(t) = \frac{aT}{\pi}\sum_{m=-\infty}^{m=\infty} J_m^2(\mathbf{GW})\cos(m\omega_S t) = \frac{aT}{\pi}J_0\{2\mathbf{GW}\sin(\omega_S t/2)\} \quad (23)$$

**2).** A smooth function of mosaic distribution and ultrasound of relatively low frequency, $f_S<<\Gamma$. $\Gamma$ is a typical frequency at which distribution function $F$ of the mosaic block orientation is changed (7a – 7d).

Using (7c) and taking into account the known relation $\sum_{m=-\infty}^{m=\infty} J_m^2(\mathbf{GW}) = 1$ [15] we find that $A(P_0) = 2aF(-P_0)$ and the low-frequency ultrasound effect on the intensity of quasi-elastic scattering *Iht* is very weak (Fig.2). We have for the gaussian distribution of mosaic block disorientation (11) $\Gamma = V_n \eta \cos\Omega_B/\lambda \sim 1$ GHz when mosaicity $\eta \sim 5$ arcmin, $\lambda = 0.58$ nm, $\Omega_B = 60°$. Therefore ultrasonic effect on the intensity of quasi-elastic scattering should be observed at the AW frequency ~ 860 MHz. However in our experiments [9] – [14] this effect was observed at the AW frequency ~70÷110 MHz (see below for the explanation).

**3)**. Thick crystal. Deeply modulated mosaic distribution function *F*, which is not smooth but contains deep narrow troughs. These troughs correspond to the absence of mosaic blocks with definite orientation in the plane of scattering. The presence of such troughs leads to a decrease in the intensity of quasi-elastic scattering ***Iht*** because for particular incident neutrons the Bragg condition isn't satisfied (when ultrasound isn't excited). However these neutrons could be inelastic scattered with absorption or emission of ultrasonic phonons. This inelastic process intensifies quasi-elastic scattering despite a week decrease in the scattering "allowed without ultrasound". It is necessary to note that the angular position of these troughs as well as their width and other parameters could differ for different parallel planes of neutron scattering (at the same reflection). In the general case such troughs are not always the same as those on the reflectivity curve corresponding to the sample. The ordinary reflectivity curve can be smooth despite of presence of the discussed troughs due to the averaging over different parallel planes of scattering. Therefore these troughs are difficult to observe by means of the diffraction of a high-energy monochromatic γ-quantum emitted from a radioactive source. Physically picture of troughs is corresponding to the presence of blocks in single crystal sample. Interesting that strong block presence was detected in the process of our KBr s.c. neutron orientation test in HMI, Berlin. We have not managed in literature any guidance as to how such troughs could be described. Therefore we propose a simple model illustrated the role of troughs. This approach consists in constructing a distribution function for mosaic blocks containing some random parameters followed by averaging the solutions of kinetic equations over these random parameters. Experimental results are fitted in frame of trough model .

Let us suppose that we have a function with gaussian distribution of mosaic block orientations:

$$f_G = \frac{1}{\sqrt{2\pi}} \exp(-x^2/2), \quad x = hx * k \qquad (24)$$

where *hx* is a small angular step; *k* is an integer. We will introduce some random value, *gg(k)* as

$$gg(k) = \sin(2\pi(\frac{k}{Nh} + Amph * (R\mathrm{mod}(k) - 0.5))) \qquad (25)$$

Here *Nh* is an integer, *0<Rmod(k)<1* is a random value homogeneously distributed within the interval 0..1; *Amph* is the depth of the random phase modulation. We have calculated the intensity of quasi-elastic

scattering and the NSE signal applying in (3) the function $f_M$ instead of $F$ as

$$f_M(k*hx) \sim \exp(-(hx*k)^2/2)(1+dm*\mathrm{sgn}(gg(k))),$$
$$\mathrm{sgn}(x>0)=1, \quad \mathrm{sgn}(x<0)=-1 \qquad (26)$$

where $dm$ is the depth of the amplitude modulation, $0 \leq dm \leq 1$. Wide class of mosaic distribution function – from gaussian (dm=0) to deeply modulated could be described by means of function $f_M$. Without random phase modulation (Amph=0) the function $f_M(x)$ contains a regular system of troughs, the width of each trough being equal to $Nh*hx/2$ (Fig.2a). An example of the randomly modulated distribution function is shown on Fig.2b and Fig.2c. As a rule, random phase modulation leads to the narrowing of troughs for the case of $Nh \gg 1$.

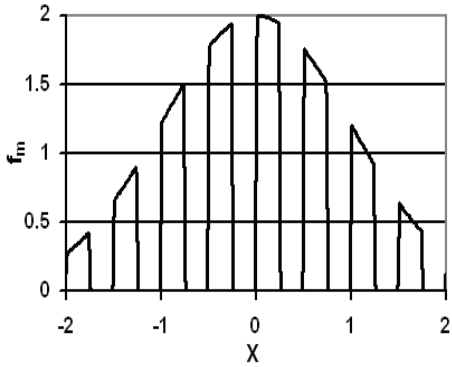

Fig. 2a) Mosaic distribution function, $f_m$ (26). Random modulation removed. *Amph*=0, *hx*=0.01, trough width *Nh*=50, depth of amplitude modulation *dm*=1.

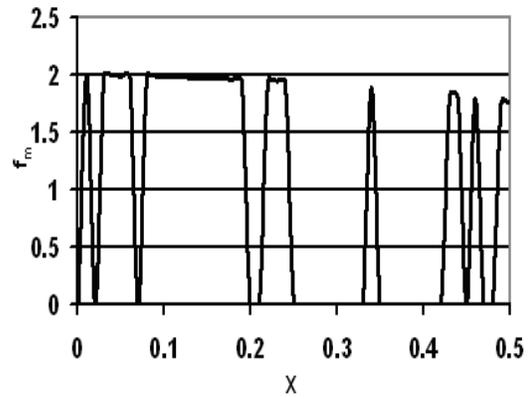

Fig. 2b) Mosaic distribution function $f_m$ (26). Random modulation. *Amph*=0.46, *hx*=0.01, trough width *Nh*=50, depth of amplitude modulation *dm*=1.

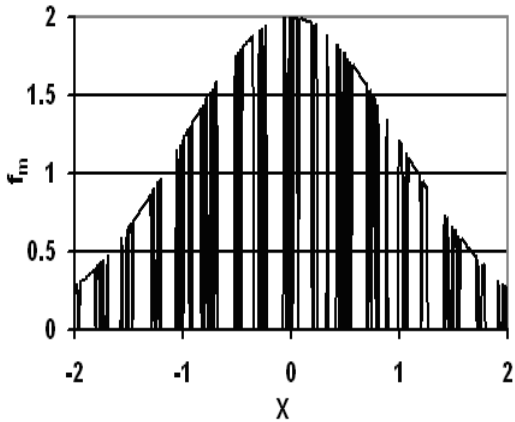

Fig.2c). Fragment of the Fig.2b.

It is obvious that we should average the solutions of the kinetic equations over varying random realizations of the mosaicity distribution function (37). The ultrasonic effect will be *lost* if we simply average modulated mosaic distribution function *inside* the kinetic equation over a random event.

The calculated results are illustrated by Fig.3a, 3b, 3c, 3d and Fig.4 for the case of coherent ultrasonic wave and the Laue scattering geometry ( Bragg scattering results are similar for the most cases).

An ultrasonic effect at the diffraction is strong for the case of higher ultrasonic frequency and deep troughs (Fig.3a).

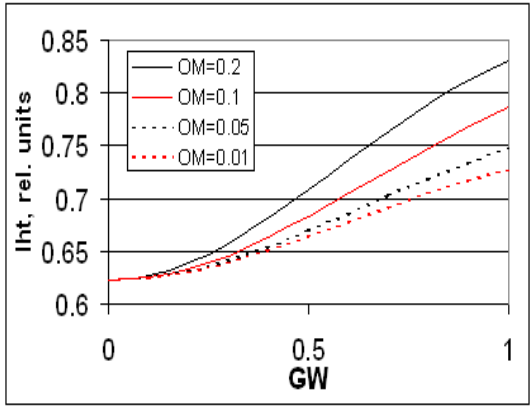

Fig. 3a) Laue geometry. The plot of calculated quasi-elastic scattering intensity, Iht, vs. ultrasonic wave amplitude. Dimensionless crystal thickness $T$=1.5, $hx$=0.01, $Nh$=50, $dm$=1.0, $Amph$=0.46. Dimensionless frequency, $OM$, is shown in the plot.

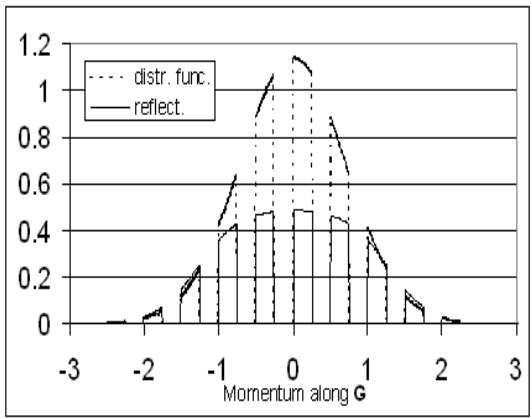

Fig. 3b) Mosaic distribution function and neutron reflectivity of a single crystal with regular (nonrandom) mosaic disorientation distribution function. Ultrasound is removed, $GW$=0, $T$=1.5, $hx$=0.01, $Nh$=50, $dm$=1.0, $Amph$=0.

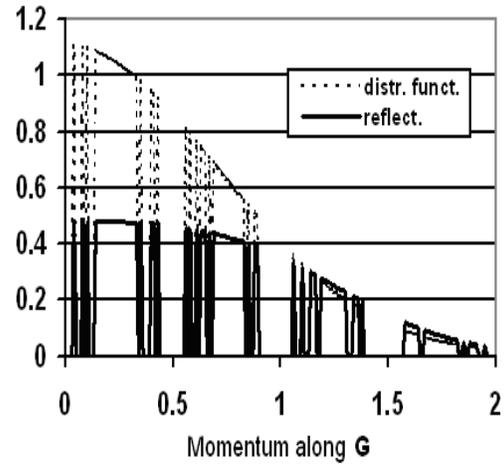

Fig. 3c) Mosaic distribution funvtion and neutron reflectivity of a single crystal with random mosaic misorientation distribution function (a curve fragment). Ultrasound is removed , $GW$=0, $T$=1.5, $hx$=0.01, $Nh$=50, $dm$=1.0, $Amph$=0.46.

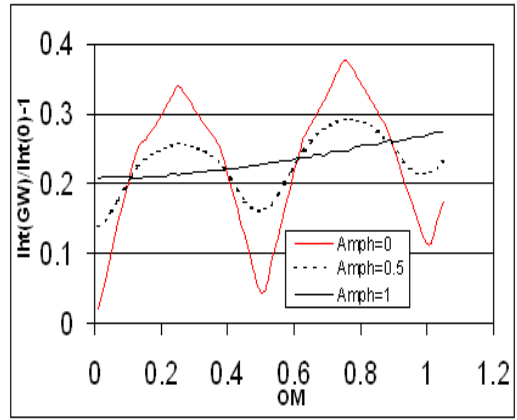

Fig. 3d) Dependence of intensity (*Iht*) vs. ultrasonic frequency *OM*. $GW$=1, $T$=1.5, $hx$=0.01, $Nh$=50, $dm$=1. The calculated values strongly depend on the parameter of random modulation, Amph, are shown in the plot.

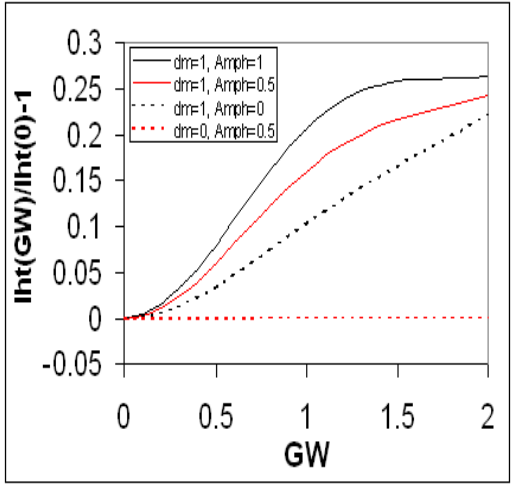

Fig. 4) Dependence of ultrasonic effect on *GW* and *Amph*; *OM*=0.05, *T*=1.5, *hx*=0.01, *Nh*=50.

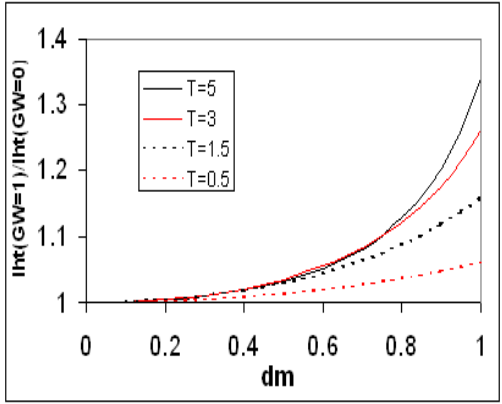

Fig. 5) Dependence of ultrasonic effect on the depth of amplitude modulation, dm, for different *T*; *OM*=0.05, *hx*=0.01, *Nh*=50, *GW*=1.0, *Amph*=0.5.

In order to understand the role of random modulation (*Amph*≠0) we will consider the neutron scattering intensity *Iht* without ultrasonic excitation of s.c. for the cases of regular (Fig.3b) and randomly modulated (Fig. 3c) mosaic distribution functions. In both the cases the central part of the *Iht* curve (not including the troughs) corresponds to the strong secondary extinction, while its sides — to the kinematic scattering; without ultrasound the values of *Iht* are almost the same in both the cases. However, the presence of numerous narrow troughs in the latter case indicates that the ultrasonic effect here will be quite different. In the former case (regular troughs) the ultrasound produces scattering of neutrons near the sides of troughs, therefore ultrasonic effect is weak at a low frequency *OM* and strong at a higher frequency. Thus one can say that ultrasound "fills" troughs in scattering. At last at the higher frequency ultrasound excites transitions mainly between the troughs and between rises in the distribution function. Therefore the ultrasonic effect on *Iht* is *decreasing*. After that the picture is more or less repeated, and we have a periodic curve for the ultrasonic effect as OM function (Fig. 3d). For the second case — *Amph*≠0 — the ultrasound fills narrow troughs during the scattering at a relatively low frequency. Therefore *Iht* is rapidly increasing at low frequencies and after that ultrasonic effect is saturated and becomes similar to that for the case of non-random modulated distribution function.

Regular mosaic distribution (*Amph*=0, trough width=0.25) leads to the existence of a periodic structure at $f_M(P)$ and *Iht(OM)* (Fig. 3b). This structure is suppressed in the case of randomly modulated mosaic structure (25), *Amph*=0.5 and *Amph*=1.

Ultrasonic effect on *Iht* is increasing at large values of *OM* (but not exceeding 0.25) and small values of *Amph*.

Randomization (increasing *Amph*) causes a strengthening of the ultrasonic effect at higher

OM values. (Fig.3c). Also, randomization leads to the appearance of narrow troughs, with the ultrasonic effect corresponding to small *GW* values increasing. The inelastic scattering is very small at *GW*<<1. Therefore in this region *Iht(GW)-Iht(GW=0)*~*(GW)*$^2$ (see Fig.4). Without troughs (the depth=0) *Iht* is independent of the ultrasound with frequency *OM*=0.03 (Fig.4). It seems reasonable to clarify the surmised role of the trough depth. The ultrasonic effect is strong and very sensitive to the trough depth for thick crystals, that is, having strong secondary extinction (Fig.5). This sensitivity is less for crystals of medium thickness (*T*=0.5). However, in both cases the presence of deep troughs is necessary for the existence of observable ultrasonic effect on the intensity of quasi-elastic scattering.

### 2.1.2 The effect of the incoherent ultrasound.

The effect of coherent and incoherent ultrasound is almost the same at relatively small ultrasonic amplitudes, $|GW|<<1$, and one-phonon scattering. It could be seen from expression (17) that

$$J_0^2(\mathbf{GW}) \approx 1 - (\mathbf{GW})^2, \quad J_1^2(\mathbf{GW}) \approx (\mathbf{GW})^2/2,$$
$$<J_0^2(\mathbf{GW})> \approx 1 - <(\mathbf{GW})^2>, \quad <J_1^2(\mathbf{GW})> \approx <(\mathbf{GW})^2>/2 \quad (27)$$

The difference between coherent and incoherent ultrasound effects could be essential for the case of multi-phonon scattering (see below).

An interesting qualitative effect could be observed with the help of NSE technique. The NSE signal *S(t)* can be negative for the case of coherent high-amplitude ultrasound. However *S(t)* remains to be non-negative for the case of incoherent high-amplitude ultrasound (Fig.6).

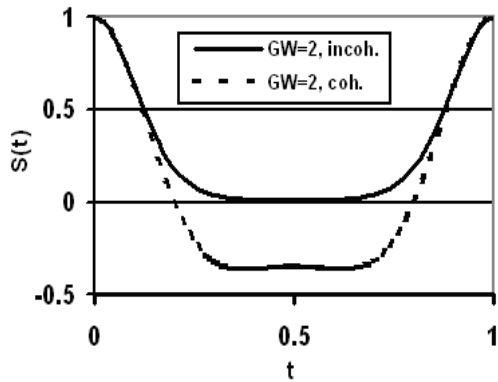

Fig. 6) Calculated NSE notmalized signal for the cases of coherent and incoherent ultrasound. NSE signal is non-negative in the last case. *OM*=0.05, *T*=1.5, *hx*=0.01, *Nh*=50, *dm*=1.0, Amph=0.5.

In our experiments [9]–[14] we observed a qualitatively similar effect with strongly ultrasonically excited silicon s.c. plate — the NSE signal dropped almost to zero but remained non-negative.

### 2.4.3 Small-sized mosaic blocks.

Parameter *A* is independent on the ultrasonic amplitude. (21), (22). Therefore the quasi-elastic scattering intensity is independent of the ultrasonic excitation; the ultrasound simply "transforms" elastic scattering to inelastic. The NSE signal *S(t)* can be negative in the case of coherent ultrasound. For incoherent ultrasound $\varphi=0$ in (22), so *S(t)* is non-negative.

## 3. Experiment

**Instruments.** To study quasi-elastic scattering ToF spectrometer V3 (NEAT) [21] was applied in a standard mode. Neutron Spin Echo measurements [2], [9] – [14] were taken on IN11, ILL spectrometer [17], [18].

**Ultrasound.** Longitudinal (LAW) and transversal (TAW) acoustic waves were excited by standard $LiNbO_3$ piezoelectric transducers (frequency 28÷111 MHz).

**Samples**. PG-graphite, KBr and artificial mica (fluorophlogopite) single crystals.

### *3.1. PG-graphite.*

NSE studies of symmetrical Bragg reflection (002) were carried out on a 70x70x6 $mm^3$ PG-graphite plate, with neutron wavelength λ=5.8 Å, LAW, f=30.5 MHz In the studies inelastic scattering was observed. However the intensity of quasi-elastic scattering, Iht, was independent of ultrasonic excitation.

### *3.2. KBr.*

### 3.2.1. ToF studies.

Symmetric Bragg reflection (200) was studied for the case of mosaic KBr single crystals (neutron wavelength $\lambda$=0.5 nm; energy resolution 85 μeV. TAW, *f*=78 MHz). The sample was fixed in a crystal holder and placed into the cryostat chamber of V3 instrument at a depth of 152 cm. The plane of reflection was located around the horizontal axis outside the cryostat. The cryostat (with the sample) was rotated around the vertical axis. It was filled by He gas at the room temperature. The gas temperature was stabilized with the accuracy of ±0.01 K. The samples was heated by ultrasound up to the temperature not greater than 50–60ºC (the heating regime was also estimated from the intensity of inelastic neutron scattering by thermal phonons). Previously the whole set-up was checked by studying the influence of LAW, *f*=70.2 MHz on the intensity of neutron diffraction (111) in the perfect silicon s.c. plate. The maximum value achieved for the ultrasonic amplitude in Si was Wmax=1.24 Å.

The ultrasonic effect on the intensity of the full Bragg's scattering and the quasi-elastic scattering is shown in Fig.7 and Fig.8, respectively.

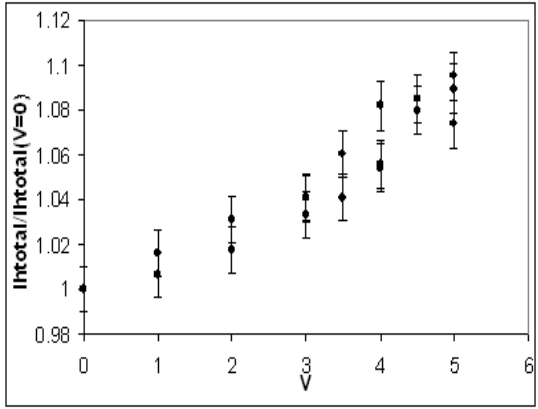 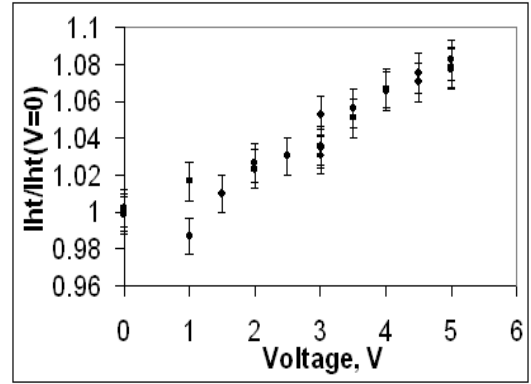

Fig. 7) KBr (200) reflection. Bragg total intensity of scattering, *Ihtotal*, vs. generator voltage, *V*; TAW, $f_e$ =78 MHz. Energy analysis is removed.

Fig. 8) KBr (200) reflection. Intensity of quasi-elastic scattering, *Iht*, vs. generator voltage *V*; TAW, $f_e$ =78 MHz. Energy resolution is 0.085 meV or 20 GHz. Contribution of scattering at the phonons with frequency large than 20 GHz is substracted.

The observed ultrasonic effect was approximately proportional to the value of (**GW**)$^2$ in the limit of low voltage across the transducer, as was expected from the theory. Our energy resolution makes it possible to subtract scattering by the phonons with frequency >20 GHz (Fig.8). However the inelastic scattering by thermal phonons in KBr heated by ultrasound is weak, therefore Fig.7 and Fig.8 are similar. The results of ToF studies lead to the following conclusion: in mosaic KBr single crystals the h. f. ultrasound increases the intensity of quasi-elastic neutron scattering by up to ~(8±1)%, and this increase is not connected with ultrasonic heating of the sample.

### 3.2.2. NSE studies.

It seems interesting to compare the results of ToF measurements with our early NSE studies of inelastic neutron scattering by ultrasound in KBr single crystals [10, 14]. We used the IN11 spin-echo spectrometer at the ILL, Grenoble. Its parameters were: $\lambda \approx 5.85$Å, $\theta_B$=62.5°; the neutron beam collimated to 0.5° FWHM and monochromatized to $\Delta\lambda/\lambda$=21%, NSE time scale 9 nsec. We excited TAW with frequency  $f$=111 MHz. and studied symmetric Bragg reflection (002) from the largest surface of a KBr single crystal. A typical observed normalized NSE signal *S(t)* is shown in Fig.9.

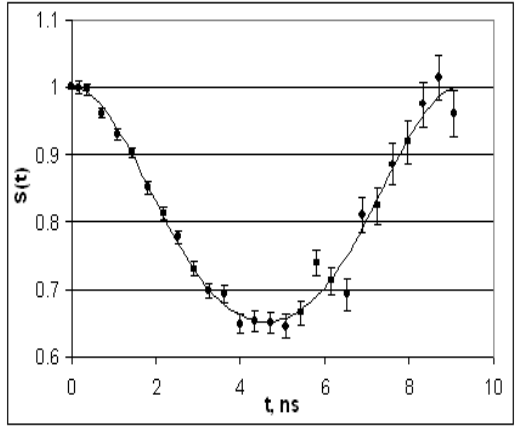

Fig. 9) KBr (200) Bragg reflection. Typical NSE normalized signal $S(t)$ as observed at moderate ultrasonic power. TAW, $f_e$ =111 MHz, $V$=30 mV — before amplifier, amplifier gain ~30. Fitting curve, $S_f(t)$ (39) .. Fitting parameters $m1$=0.1738±0.0044, $m2$=0.0165±0.0048, $m3$=0.1084±0.0011 GHz, Chisq=0.0098751, factor R=0.989.

NSE signal was approximated by the following function (see Fig.9):

$$S_f(t) = 1 - m1 - m2 + m1\cos(2\pi t m3) + m2\cos(4\pi t m3) \quad (28)$$

where m1, m2 are amplitude and m3 frequency fitting parameters. The relative values of elastic, $Ir(0)$, one-phonon, $Ir(1)$, and double-phonon scattering, $Ir(2)$, can be written as follows.

$$Ir(0) = 1 - m1 - m2, \quad Ir(1) = m1,$$
$$Ir(2) = m2$$

The absolute values of the corresponding scattering will be

$$Ih(0) = Iht(1 - m1 - m2),$$
$$Ih(1) = Ihtm1, \quad Ih(2) = Ihtm2$$

Here $Iht$ is the total intensity of quasi-elastic scattering (depending on the ultrasonic wave amplitude for the case of mosaic s.c. – see Figs. (8), (10). The intensities of elastic and one-phonon scattering are high: m1=0.1738±0.0044, while the intensity of double-phonon scattering is low, therefore the corresponding relative error is large: m2=0.0165± 0.005. The value m3 is the fitting parameter of frequency: m3=0.1084±0.0011 GHz.

It is interesting that the ultrasonic frequency in the NES observations ($f_{NSE}$=108.4÷1.0 MHz) was measured with a small error; this frequency was only slightly lower than that of electronic excitation ($f_e$=111MHz). The dependences of quasi-elastic, elastic, one- and double-phonon scattering intensities on the voltage across the transducer are shown in Figs.10, 11.

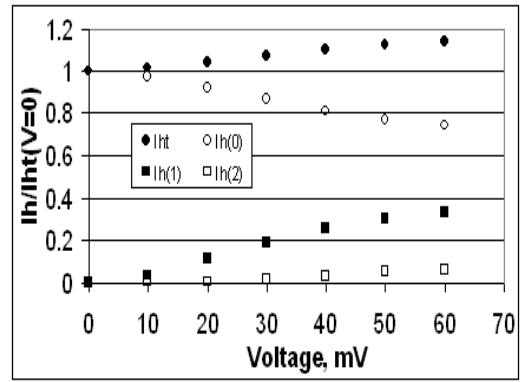

Fig. 10) Spectral distribution of neutron scattering by ultrasonically excited KBr s.c. Intensity of quasi-elastic Iht, elastic Ih(0), one- Ih(1) and double- Ih(2) phonon scattering vs. generator voltage.

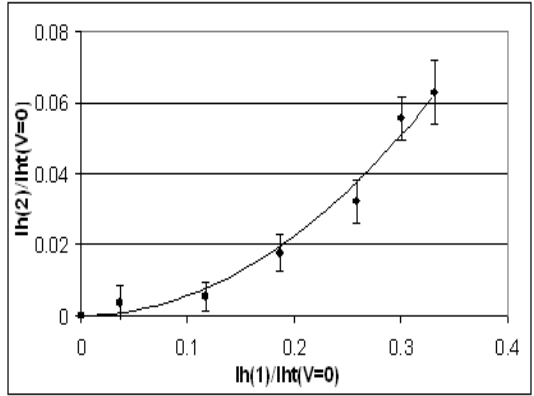

Fig. 11) Double-phonon scattering intensity, $Ih(2)$, vs. one-phonon scattering, $Ih(1)$, in KBr (experimental points and fit $y=0.565x^2$).

In agreement with theory the observed (Fig.10) ultrasonic effect on $Iht$, $Ih(0)$ and $Ih(1)$ is proportional to the value $(GW)^2$ (more exactly to $V^2$) at a low voltage V across the transducer. Also, we in agreement with theory we had $Ih(2) \sim Ih(1)^2$ in the region of low ultrasonic amplitude (Fig.11).

We observe that ultrasound increases the total intensity of scattering in KBr by 14% (Fig.10). The KBr samples for ToF and NSE experiments as well as the conditions of ultrasound excitation were different. Therefore we could expect (and observed) that there is only qualitative agreement between ToF and NSE results. The energy resolution of 85μeV obtained in ToF experiment corresponds to the frequency $\sim 2 \times 10^{10}$ Hz and can be improved. Some information about the contribution of very h.f. phonons probably could be derived from the observed NSE signal $S(t)$. The second point on the S(t) curve of Fig. 9 corresponds to $t=0.18$ ns. We should have a sharp decrease in the NSE signal within this short time interval if phonons with frequency $\geq 3$ GHz were important in the scattering. However we did not notice any sharp changes on the $S(t)$ curve at $t \leq 0.18$ ns.

The results of ToF and NSE experiments lead to the same conclusion: the h.f. ultrasound increases the intensity of quasi-elastic neutron scattering in mosaic KBr single crystals. This increase does not depend on the ultrasonic heating of the sample.

### 3.3. Mica.

At last qualitatively similar results were obtained [12] for the case of neutron scattering by ultrasonically excited single crystal plates of artificial mica (fluorophlogopite), with sizes 50x25x(1-2) mm³ and 25x25x1 mm³. In our experiment we studied symmetrical Bragg's reflection (200). The LAW was excited by a $LiNbO_3$ transducer. Since very cold neutrons were applied ($\lambda=11.02$ Å, $d \approx 9.9$Å, $\theta_B \approx 33.47°$), for NSE Fourier time the value of $t \sim 63$ ns was achieved. The high-quality NSE signal in mica is shown in Fig.12.

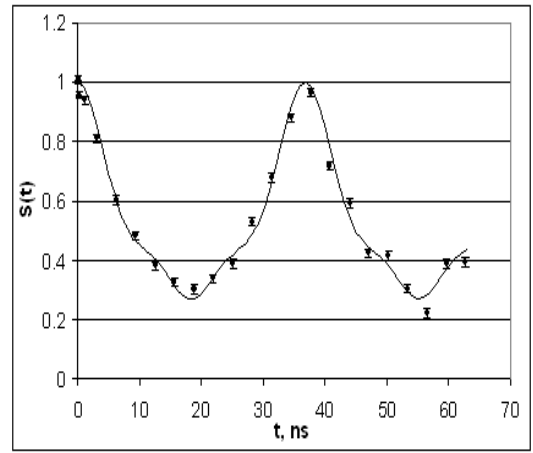

Fig. 12) Mica (200) Bragg reflection.. NSE normalized signal $S(t)$. LAW, $f_e = 28$ MHz, $V$=150 mV. Fitting curve, $S_f(t)$,

$$S_f(t) = 1 - m1 - m2 - m3 + m1\cos(2\pi tm4)$$
$$+ m2\cos(4\pi tm4) + m3\cos(6\pi tm4)$$

.Fitting parameters $m1=0.3101\pm0.0094$, $m2=0.0838\pm0.0092$, $m3=0.0551\pm0.0096$, $m4=0.02716\pm0.00015$ GHz, Chisq=0.026, factor R=0.99.

We took into account elastic, one-, double- and even three-phonon scattering in the fitting function.

The frequency in NSE observations was $f_{NSE}=27.16$ MHz, which is by 3% smaller than the generator frequency of 28 MHz.

Similar fitting allowed us to find the dependences of quasi elastic, $Iht$, elastic, $Ih(0)$, one-phonon, $Ih(1)$, double-phonon, $Ih(2)$, and even triple-phonon, $Ih(3)$, scattering intensities on the generator voltage. The ultrasound strongly (~50%) increases the intensity of quasi-elastic scattering in mica (Fig. 13). We have not found any serious contribution (greater than 5% at $t<0.2$ ns) of thermal phonons in this scattering.

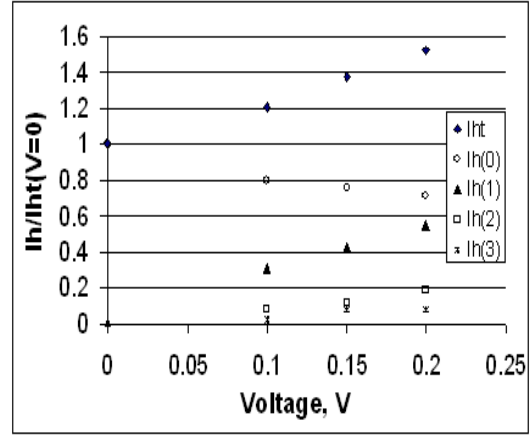

Fig. 13) Spectral distribution of neutron scattering by ultrasonically excited mica s.c. LAW, $f_e = 28$ MHz.

The qualitative picture of ultrasonic effect in mica is similar to that in KBr s.c. Unfortunately, the number of experimental points is insufficient, especially at low transducer voltage (Fig.13). Therefore we will mainly restrict ourselves to the analysis of KBr experimental results given below.

## 4. Analysis of experimental results

### 4.1. PG-graphite

While observing inelastic neutron scattering by ultrasound we found that the intensity of quasi-elastic scattering, $Iht$, did not depend on the conditions of ultrasonic excitation. Taking into account very large imperfection of PG single crystals, we may derive a conclusion that the ultrasonic effect can only be considered for small-sized blocks (see the discussion after (2)).

### 4.2. KBr

We will concentrate on the analysis of Bragg's geometry reflection (200) data for KBr single crystals using the following parameters: neutron wavelength $\lambda=0.585$ nm, KBr lattice constant $a=0.6598$ nm, scattering length $a_K=3.67$ fm, $a_{Br}=6.795$ fm, $\Theta_B=62.5°$, ultrasound frequency $f=111$MHz, and TAW wave length $\lambda_S=12.3$ μm. The extinction length, $\tau$, and parameter Q [16] are written as follows.

$$\tau = \pi\cos(\theta_B)/(N\lambda|F|) \approx 17\mu m, \quad Q = \lambda^3 N_c^2 F^2/\sin(2\theta_B), \quad Q/\eta/\sqrt{2\pi} \approx 71.3/\eta_{arc\min} \quad (29)$$

Here $F$ is a structural factor, $F=41.86$ fm, $N_c$ is the density of cubic elementary cells, $N_c=3.48*10^{21}$cm$^{-3}$, $\eta$ is the mosaicity disorientation. The width of Bragg's plateau is $2\delta\theta_0=\lambda/\tau/\sin(\theta_B)\approx 8$ arcs. Let us estimate the value of ultrasonic frequency $f_t$ corresponding to the transition between distribution functions (26) of central and lateral mosaic disorientations. The corresponding exponential factor in (26) is equal to the value of exp(-0.5) if $f_t=\eta V\cos(\theta_B)/\lambda=156$ MHz/arcmin$\times\eta_{arcmin}$. Now we should estimate the value of $\eta$. The standard equations of secondary extinction are written in the form [16]:

$$\partial I_0/\partial z = -\tau_1(I_0-I_h),..., \quad \tau_1 = QW_1(\theta-\theta_B)/\sin(\theta_B), \quad W_1(\Delta) = \frac{1}{\eta\sqrt{2\pi}}\exp(-0.5(\Delta/\eta)^2)$$

Here $I_0$, $I_h$ is the intensity of incident and diffracted beams, respectively.

We have done calculations for the case of symmetric Bragg's scattering with deeply modulated distribution function $f_M$ (25) and incoherent ultrasonic wave. The optimal dimensionless fitting parameters (25) are the following: ultrasonic frequency OM=0.05, thickness T=5.0, momentum step hx=0.01, Nh=50, modulation depth of the distribution function dm=0.95, Amph=0.460.

We have found from fitting calculations that $QW(0)T_{Lab}/\sin(\theta_B)\sim 2\div3$, which corresponds to $\eta\sim(8\div20)$arcmin. Therefore the value of $f_t$ is very large, $f_t=(1.2\div3.1$ GHz However, our distribution function is deeply (95%) modulated, and the trough width hx×Nh/2 correlates with the frequency interval of (300–775) MHz. Therefore TAW with frequency OM=0.05 (that is (60–155) MHz) intensifies quasi-elastic scattering. We have calculated the energy spectrum of the diffracted neutron beam for the case of coherent and incoherent ultrasonic waves (Fig.14). The value of $|GW|$ parameter is not large in our case (<1).

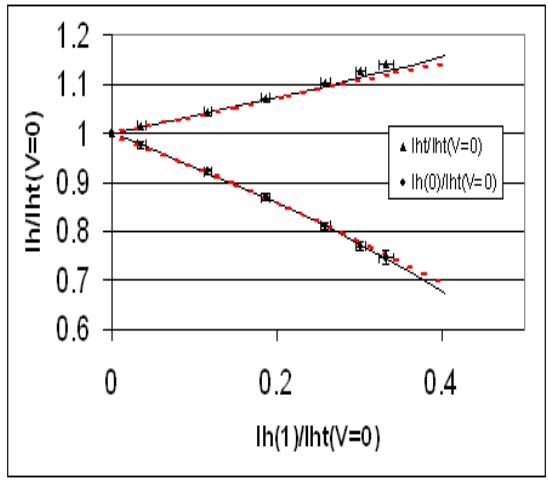

Parametric dependence of quasi-elastic and elastic scattering intensity vs. one-phonon scattering intensity. Points — experiment, solid lines — calculation for the case of incoherent ultrasound; dotted lines — calculation for the case of coherent ultrasound. Random mosaic disorientation distribution function. Dimensionless parameters: $OM$=0.05, $hx$=0.01, $Nh$=50, trough deep $dm$=0.95, $Amph$=0.46.

Therefore experimental results for $Iht$, $Ih(0)$ are well described by coherent and incoherent ultrasound approximations (Fig.14).

Fig.14) Ultrasonically excited KBr s.c. Bragg geometry, (200) reflection.

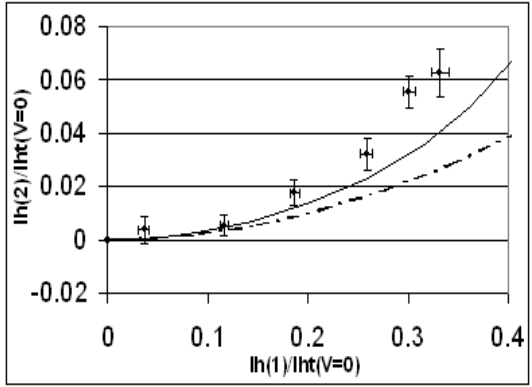

Fig. 15) Double-phonon scattering in KBr s.c. Points — experiment, solid lines — calculation for the case of incoherent ultrasound; dotted lines — calculation for the case of coherent ultrasound. Random mosaic disorientation distribution function. Dimensionless parameters: $OM=0.05$, $hx=0.01$, $Nh=50$, $dm=0.95$, $Amph=0.46$.

However double-phonon scattering intensity $Ih(2)$ is better described by the incoherent ultrasound field approximation (Fig.15). This can be explained as follows. The value of $Ih(2)$ is small, $Ih(2) \sim (GW)^4$ (roughly). It is easy to calculate, taking into account, that for the case of incoherent ultrasound $<(GW)^4> = 2<(GW)^2>^2$. In the case of coherent ultrasound $(GW)^4 = ((GW)^2)^2$. Therefore higher-order momentum $|GW|$ is larger in the incoherent case, the double-phonon scattering is stronger and better correlates with our experimental results.

Correlation between the generator voltage, V, and the calculated $|GW|$ is shown in Fig.16: $V=66*|GW|$ mV, $W \approx 0.24$ Å at a generator voltage 30 mV.

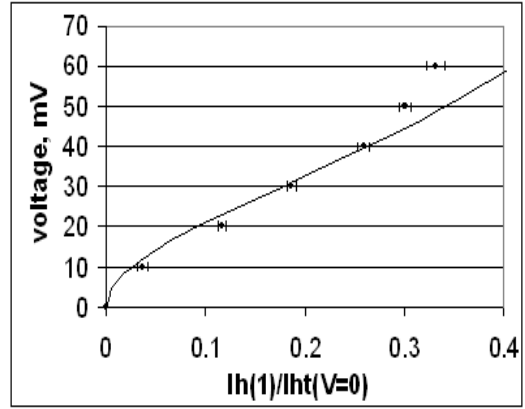

Fig. 16) One-phonon scattering in KBr. Points — experiment, solid?lines — calculation for the case of incoherent ultrasound under assumption that $V=C*GW$, fitting parameter $C=66$ mV. Random mosaic disorientation distribution function. Dimensionless parameters: $OM=0.05$, $hx=0.01$, $Nh=50$, $dm=0.95$, $Amph=0.46$.

The corresponding deformation in the acoustical wave $2\pi W/(c_s f_s) \approx 8*10^{-6}$ (velocity of TAW $V_S=2*10^5$ cm/sec). The linear correlation between V and $|GW|$ is violated at a higher voltage, ~50÷60 mV. The value of $|GW|$ is increasing *more slowly* than voltage. It is easy to show that similar violation of the linear relation between values of $|W|$ and voltage occurs if we apply Abraham's expression [20] for fitting the Ruby and Bolef Mossbauer effect data [19]. Probably, the model of incoherent ultrasound could be considered as a reasonable though limited approximation to the real multimode acoustic field description.

An example of good correlation between the observed and the calculated NSE signals is given in Fig.17.

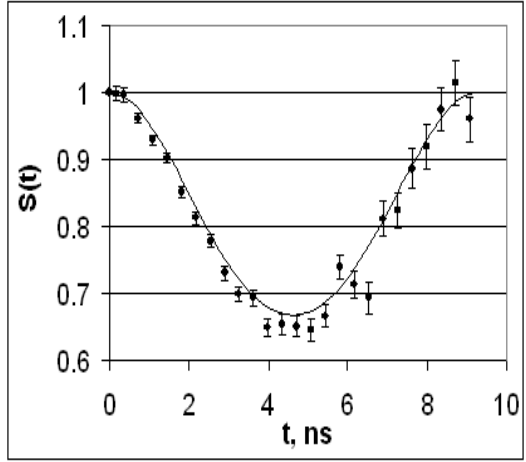

Fig. 17) KBr (200) Bragg reflection. NSE normalized signal S(t) as observed at moderate ultrasonic power. TAW, $f_e$ = 111 MHz, $V$=30 mV. Points — experiment, solid line — calculation for the case of incoherent ultrasound, $GW$=0.45. Time-scale is the same as in Fig. 9. Random mosaic disorientation distribution function. Dimensionless parameters: $OM$=0.05, $hx$=0.01, $Nh$=50, $dm$=0.95, $Amph$=0.46.

### *4.3. Mica*

It is shown above that the "deep trough" mechanism allows one to understand the ultrasonic effect at the cold neutron scattering in mosaic KBr s.c. Probably, the same approach could be applied to the case of neutron diffraction in mica (Figs.12, 13). Moreover, the increase in the intensity of quasi-elastic scattering *Iht* is even more expressed for mica than for KBr. Rectilinear dependences of *Ih(1)* and even of *Ih(2)* on the generator voltage *V* could be considered as evidence of the presence of strong secondary extinction effects. However $Ih(3) \sim V^2$. The existence of a noticeable secondary extinction effect in mica s.c. was stressed by other authors (see, e.g. [22], [23]).

Unfortunately, we have no NSE experimental data for the mode of relatively low voltage *V* (see Fig.13). More detailed comparison of our model with experiment could be done after additional NSE measurements at small *V* values.

### 5. Summary

We have applied Time-of-Flight and Neutron-Spin-Echo techniques for studying the ultrasonic effect at the cold neutron diffraction in mosaic single crystals. We found that ultrasound increases the intensity of quasi-elastic neutron scattering, *Ih*, at the ultrasonic frequency 28÷111 MHz, and this effect cannot be interpreted as a result of neutron scattering by thermal phonons arising under ultrasonic heating of the sample.

The theory of the effect is based on the supposition that the exchange of energy between a cold (slow) neutron and the ultrasound leads to the existence of a separate inelastic channel of scattering in each medium-sized mosaic block. The corresponding kinetic equations, taking into account *all* ultrasonic transitions, are formulated and *exactly* solved for the case of *arbitrary* angular distribution function *F* describing mosaic block orientation, and coherent and incoherent ultrasonic waves. Solution of these kinetic equations clearly shows that the observed ultrasonic effect on *Ih* cannot be explained under the assumption that function *F* is ordinary and smooth (e.g. a gaussian function).

We therefore propose a new model in which the distribution function $F$ is not smooth but contains deep narrow troughs. These troughs correspond to the absence of mosaic blocks with definite orientation in the plane of scattering. The presence of such troughs leads to a decrease in the intensity $Iht$ of the reflected beam without ultrasound because for particular incident neutrons the Bragg condition will not be satisfied. However these neutrons could be inelastic scattered, with absorption or emission of ultrasonic phonons. This inelastic process increases the intensity of the quasi-elastic scattering. It is necessary to note that the angular position of these troughs, their width and other parameters could differ for different parallel planes of neutron scattering (at the same reflection). In the general case such troughs are not always identical to those on the conventional reflectivity curve corresponding to the whole sample. The conventional reflectivity curve can be smooth in spite of the presence of troughs due to their averaging over different planes of scattering. Probably, these "hidden" troughs are not easy to observe by diffraction of a high-energy monochromatic γ-quantum emitted from a radioactive source. The general effect of the troughs consists in decreasing the intensity of the diffracted beam without ultrasound. We have not found in literature any description of such troughs; besides, these troughs can be described differently for different materials. Therefore we have proposed a simple approach (described above) which can illustrate the role of troughs in our case. This simple model made it possible to successfully compare our experimental data concerning not only $Iht$, but also NSE spectrum in KBr s.c., with theory.

The above theory demonstrates — in agreement with experiment — that this ultrasonic effect strongly depends on the parameters of the mosaicity distribution function. For example, the described behavior of quasi-elastic scattering intensity, $Iht$, is independent of ultrasound in PG-graphite s.c., probably due to the small size of mosaic blocks.

Of course, our model is not completely unique in details. However, model is definitely in the main - mosaic distribution function should contain deep sharp narrow troughs. Troughs parameters are defined from comparison not only with $Iht$- voltage,$V$, curve, but zero-phonon $Ih0$ -$V$, one-phonon $Ih1$ -$V$ and partly $Ih2$-$V$. It's happened because we have Neutron Spin Echo research data, that is energy spectrum. It isn't so easy to consent model with experimental data $Iht – V, Ih0 - V, Ih1 -V$, model parameters are not so free as could be suppose for the first look. It seems that our model is also natural for crystal contained large blocks that are not smoothly oriented for respect each other.

In general, this ultrasonic effect could be considered as some test of "perfection" of a mosaic structure.

It seems reasonably to apply our approach to the research of ultrasonic effect at the neutron diffraction in different technologically important mosaic s.c. (plastically deformed Ge and others). The roles of ToF and NSE techniques are different in similar studies. The former allows for more or less fast observations of ultrasonic effect on the intensity of quasi-elastic scattering at various neutron wave length. The latter (NSE method) is complicated but it allows for studying not only $Iht$ but also the whole quasi-elastic energy spectrum of the diffracted beam (ultrasonic satellites included). At last we could note that serious increasing of the intensity of quasi-elastic scattering created by ultrasound in mica (Fig.13) could be applied (in general) for the optimization parameters of neutron scattering devices used artificial mica.

It is known that attempts to grow magnetic single crystals with the "perfection" comparable with the quality of silicon or quartz s.c. have not been successful so far. We believe that our approach can be applied in the investigations into ultrasonic effects at the diffraction in magnetic s.c.

In seems reasonable to stress some important peculiarities of our NSE measurements.

1). **KBr,** (Fig.9). NSE time $t$=9 ns; *fitting* frequency (energy) error $\delta f$=±1.07 MHz (4 neV); phase error $\delta\varphi=2\pi t\delta f \approx$ ±0.06. The formal 'standard' NSE resolution $\delta f_S$=1/(2π $t$) ≈ 17.6 MHz.

2). **Mica,** (Fig.12). NSE time $t$=63 ns; $\delta f$=±28*0.0055≈±0.15 MHz (0.6 neV); $\delta\varphi=2\pi t\delta f \approx$ ±0.06. 'Standard' NSE resolution $\delta f_S$=1/(2π t) ≈ 2.5 MHz.

As is seen the frequency fitting resolution, $|\delta f|<<\delta f_S$, is very high in our experiments. Neutron time of flight through the IN11 spectrometer is ~5 ms. That means that we had a classical working regime, far from the quantum limit ~1/5ms=200Hz. 'Standard' NSE resolution $\delta f_S$ is applicable when we have no *a priori* information about the NSE signal. It does not relate to our case because we excite ultrasound by an external harmonic generator. In general, cosine type signal *S(t)* could be reconstructed if we had highly precise data in a finite time interval. The high quality of fitting (Figs. 9, 12) can be considered as evidence that the assumption $S_f(t)$ (28) about the shape of NSE signal is correct and that the NSE spectrometer ( IN11, ILL) used by us is an instrument of high stability.

It is important that we have obtained such a small value of $\delta f$ only for the case of h.f. signal. This could be explained as follows. Let us suppose that we want to derive ultrasonic frequency $\omega_l$ from the value of NSE signal $\sim\cos(\omega_l t)\approx 1-(\omega_l t)^2/2$ at the mode $\omega_l t<<1$. In this case we should have counted $\sim 1/(\omega_l t)^4$. For the case of a high frequency NSE signal, $\omega_{hf} t\sim\pi$, we should have a much smaller value of count $\sim 1/(\Delta\omega t)^2$ in order to find $\omega_{hf}$ with the error not exceeding $\sim\pm\Delta\omega$ ($\Delta\omega t<<1$). Therefore in our experiments we measured ultrasonic frequency ~111 MHz with accuracy ~ several MHz but could not study ultrasonic signal with frequency ~1÷3 MHz by means of the same experimental equipment!

## Acknowledgments

The authors are deeply grateful to E. Raitman, H. Bordallo, A. Snigirev, K. Herwig, M.L. Crow, P. Allenspach, K. Prokes, and V. Gavrilov for their important assistance, to S. Silstein, V. Somenkov for very interesting discussions. We would like to thank also the technical staff of BENSC and ILL for their great help and support during our experiments. This research project has been supported by the European Commission under the 6th Framework Program through the Key Action: Strengthening the European Research Area, Research Infrastructures. Contract n[0]: HII3-CT-2003-505925 (NMI3).